\documentclass[a4paper,12pt]{article}
\usepackage[top=1.25in, bottom=1.25in, left=1.05in, right=1.05in]{geometry}
\usepackage[utf8]{inputenc}
\usepackage[T1]{fontenc}
\usepackage[english]{babel}

\usepackage{natbib}
\usepackage[section]{placeins}
\usepackage{setspace}
\usepackage{enumitem}
\setlist{noitemsep}  % Reduce space between list items (itemize, enumerate, etc.)

\usepackage{regexpatch}

\usepackage{amscd,amsmath,amssymb,amsfonts,amsthm,bbm,bm,latexsym,mathrsfs,mathtools,bigints}
\usepackage[subnum]{cases}	% multi-case equations with separate number for each
\usepackage{dsfont}
\makeatletter
\newcommand*{\distas}[1]{\mathbin{\overset{#1}{\kern\z@\sim}}}	
\makeatother
\newcommand*\abs[1]{\left|#1\right|}		% absolute value
\newcommand{\norm}[1]{\left\lVert#1\right\rVert} 		% norm

\allowdisplaybreaks

\theoremstyle{remark}

\theoremstyle{plain}

\usepackage{algorithm, algorithmic}

\RequirePackage[colorlinks=true, citecolor=blue, urlcolor=blue]{hyperref}

\usepackage{xcolor,epsfig,epstopdf,rotating,graphics,graphicx, subcaption} %subfigure 
\usepackage[width=0.9\textwidth, font={footnotesize}]{caption}

\usepackage{rotating,lscape,pdflscape}
\usepackage{booktabs,longtable,float,array,multirow,colortbl,adjustbox}
\newcolumntype{C}[1]{>{\centering\arraybackslash}p{#1}}
\definecolor{dgray}{gray}{0.65}
\definecolor{lgray}{gray}{0.90}
\definecolor{bostonred}{rgb}{0.8, 0.0, 0.0}
\definecolor{candyapplered}{rgb}{1.0, 0.03, 0.0}
\definecolor{ferrarired}{rgb}{1.0, 0.11, 0.0}
\colorlet{lred}{candyapplered!30!white}
\colorlet{dred}{ferrarired!50!lred}
\colorlet{dred}{dgray!90!white}
\colorlet{lred}{lgray!40!white}

\usepackage{pifont}

\usepackage{physics}
\usepackage{comment}

\usepackage{epsfig}

\def\I {\mathbb{I}}
\def\R {\mathbb{R}}
\def\d {\mathrm{d}}
\def\bA {\mathbf{A}}
\def\bB {\mathbf{B}}

\def\bc {\mathbf{c}}
\def\be {\mathbf{e}}
\def\bx {\mathbf{x}}
\def\by {\mathbf{y}}

\def\bY {\mathbf{Y}}
\def\bX {\mathbf{X}}
\def\bC {\mathbf{C}}
\def\bE {\mathbf{E}}
\def\bG {\mathbf{G}}
\def\bF {\mathbf{F}}
\def\bV {\mathbf{V}}
\def\bU {\mathbf{U}}
\def\bM {\mathbf{M}}
\def\bI {\mathbf{I}}
\def\bzero {\mathbf{0}}
\def\bSigma {\mathbf{\Sigma}}
\def\balpha {\boldsymbol{\alpha}}
\def\bbeta {\boldsymbol{\beta}}
\def\bdelta {\boldsymbol{\delta}}
\def\bgamma {\boldsymbol{\gamma}}

\def\bmu {\boldsymbol{\mu}}

\def\bmuo {\overline{\boldsymbol{\mu}}}
\def\Lambdau {\underline{\smash{\boldsymbol{\Lambda}}}}
\def\Lambdao {\overline{\boldsymbol{\Lambda}}}

\def\nuu {\underline{\smash{\nu}}}
\def\nuo {\overline{\nu}}
\def\Psiu {\underline{\smash{\boldsymbol{\Psi}}}}
\def\Psio {\overline{\boldsymbol{\Psi}}}

\def\rmax {r_{\max}}
\def\qg {q_{\gamma}}
\def\rank {\operatorname{rank}}
\def\vec {\operatorname{vec}}
\def\cov {\operatorname{cov}}
\def\nbd {\operatorname{nbd}}

\title{\vspace{-60pt} \textbf{Bayesian Partial Reduced-Rank Regression}
}

\author{
Maria F. Pintado\thanks{Queen Mary University of London, United Kingdom,  {\color{blue}\texttt{m.f.pintadoserrano@qmul.ac.uk}}, supported by CONAHCyT (Mexico), grant no. 2021-000007-01EXTF-00090.}
\and
Matteo Iacopini\thanks{Queen Mary University of London, United Kingdom, \color{blue}\texttt{m.iacopini@qmul.ac.uk}}
\and
Luca Rossini\thanks{University of Milan, Italy and Fondazione Eni Enrico Mattei, \color{blue}\texttt{luca.rossini@unimi.it}}
\and
Alexander Y. Shestopaloff\thanks{Queen Mary University of London, United Kingdom and Memorial University of Newfoundland, Canada, \color{blue}\texttt{a.shestopaloff@qmul.ac.uk}}
}

\date{\today}

\begin{document}

\maketitle

\begin{abstract}
Reduced-rank (RR) regression may be interpreted as a dimensionality reduction technique able to reveal complex relationships among the data parsimoniously.
However, RR regression models typically overlook any potential group structure among the responses by assuming a low-rank structure on the coefficient matrix.
To address this limitation, a Bayesian Partial RR (BPRR) regression is exploited, where the response vector and the coefficient matrix are partitioned into low- and full-rank sub-groups. As opposed to the literature, which assumes known group structure and rank, a novel strategy is introduced that treats them as unknown parameters to be estimated.

The main contribution is two-fold: an approach to infer the low- and full-rank group memberships from the data is proposed, and then, conditionally on this allocation, the corresponding (reduced) rank is estimated. Both steps are carried out in a Bayesian approach, allowing for full uncertainty quantification and based on a partially collapsed Gibbs sampler. It relies on a Laplace approximation of the marginal likelihood and the Metropolized Shotgun Stochastic Search to estimate the group allocation efficiently.
Applications to synthetic and real-world data reveal the potential of the proposed method to reveal hidden structures in the data.

 \vspace*{3pt}
\end{abstract}

\section{Introduction}
\label{sec:intro}

Intrinsic group structures are prevalent in certain data types, particularly in high-dimensional datasets. These structures signify the presence of correlations among variables within these groups, and ignoring them can lead to an inefficient use of the available data. 
In the context of multivariate response-predictor analysis, a commonly adopted strategy is to perform covariate selection for each response variable, which addresses the predictor group structure \citep[see][for a review of such methods]{buch2022systematic}.
However, the group structure among response variables is typically overlooked. 

The reduced-rank (RR) regression model \citep{anderson1951estimating, izenman1975reduced, reinsel2022book} offers a more natural means of handling block structures and dimension reduction, achieved by imposing a lower-rank constraint on the matrix of coefficients, $\bC$. This assumption translates into having a smaller number of relevant linear combinations of the predictor variables as latent factors that explain the variation in all the response variables. 
Different variants of the reduced-rank regression model have been explored in the literature. 
For instance, \cite{anderson1951estimating} examines a partitioned coefficient matrix associated with a low-rank group and a full-rank group in the covariates using a predefined grouping. 
This result is further elaborated upon by \cite{velu1991reduced}, who extends it into two sets of regressors characterised by low-rank structures. 
Recently, \cite{li2019integrative} investigated an integrative reduced-rank regression model for analysing multi-view data, where each view consists of several predictors and has its own low-rank coefficient matrix. 
On a different direction, \cite{chen2012sparse} proposed a sparse reduced-rank regression that introduces row-wise sparsity in $\bC$, enabling predictors with no association to latent factors.
\cite{kim2023integrative} combined the latter two approaches in a reduced-rank regression setting with multi-source data, where each predictor set is associated with a low-rank coefficient matrix while simultaneously allowing for sparsity in both covariates and responses.

We explore an alternative generalisation of the standard RR regression model, where the reduced-rank coefficient structure applies to only an unknown subset of the response variables. In contrast, the remaining subset maintains a full-rank coefficient submatrix.
This more flexible approach has been called partially reduced-rank (PRR) regression by \cite{reinsel2006partially}. In this scenario, the set of response variables is divided into two subsets $\bY_1$ and $\bY_2$, and the reduced-rank structure is imposed on a submatrix $\bC_1$ of $\bC$, driving the relationship between $\bY_1$ and the covariates, $\bX$. The reduced-rank assumption on $\bC_1$ implies that the regression of $\bY_1$ on $\bX$ is influenced by only a limited number of predictive variables constructed as linear combinations of $\bX$.

Considering group structures within the response variables enhances our comprehension of the data in diverse fields such as macroeconomics \citep{reinsel2006partially} and genetics \citep{li2015multivariate,luo2019feature}. 
The PRR model, incorporating the proposed response groups, has a potential utility in multi-view data scenarios that consider two views in the responses, thereby enhancing model fit through the specialised structure in $\bC$.
Additionally, PRR adds flexibility to the model in accommodating complex relationships observed in real-world datasets, such as those found in economic contexts.

To the best of our knowledge, the literature on PRR regression summarised in \cite{reinsel2006partially} has always relied on an \textit{a priori fixed} grouping structure, justified by application-specific theoretical considerations or the researcher's intuition. We argue that this represents a possible reason hindering the large applicability of PRR regression in several real-world problems.
To overcome these issues, we propose the first Bayesian approach to PRR regression, where the low-rank and full-rank response groupings are unknown and directly inferred from the data. This approach considers an agnostic position about the optimal allocation of response variables and opens the possibility of using PRR models even in the absence of strong and reliable information to dictate the grouping.
Moreover, we design and implement a partially collapsed Gibbs sampler that, at each iteration, first samples both the grouping structure and the (reduced) rank relying on the Laplace method, then the remaining parameters in subsequent basic steps. We call the proposed method, Bayesian partial reduced-rank (BPRR) regression and we compare its performance against well-known specifications available in the literature. 

A simulation study underscores the strong performance of our proposed model across various scenarios. It yields adequate estimates of the low-rank group allocation and rank and achieves a lower error compared to benchmark methods in linear regression such as the pre-specified PRR or the full-rank model. The effectiveness of the model is further demonstrated in an application to macroeconomic data from the United States, which provides evidence of a significant shift in the response grouping (and rank) following the COVID-19 pandemic, both in terms of point estimates and the associated uncertainty.

The remainder of the article is as follows. Section~\ref{sec:model} briefly presents the PRR framework and describes the structure of the prior distributions.
Then, Section~\ref{sec:posteriors} describes in detail the challenges encountered in the design of the algorithm to perform posterior sampling together with the proposed solutions.
The performance of the algorithm is tested on synthetic data in Section~\ref{sec:sim}, and then it is applied to a real-world dataset about macroeconomics in Section~\ref{sec:app}.
 Section~\ref{sec:conc} draws the conclusions.

%%%%%%%%%%%%%%%%%%%%%%%%%%%%%%%%%%%%%%%%%%%%%%%%%%%%%%%%%%%%%%%%%%%%%%%%%%

\section{Partial reduced-rank regression model}
\label{sec:model}

Let $\bY \in \R^{n\times q}$ be the matrix of responses with the $i$th row denoted as $\by_{(i)}'$, $\bX\in\R^{n\times p}$ the matrix of explanatory variables with the $i$th row as $\bx_{(i)}'$, and $\bE \in \R^{n\times q}$ the matrix of innovation terms with $\be_{(i)}$ as the noise vector associated to the $i$th observation. The multivariate linear regression model is defined as
\begin{equation}
\label{eq:linearmodel}
    \bY = \bX\bC+\bE, \qquad \bE=(\be_{(1)},\ldots,\be_{(n)})', \qquad \be_{(i)} \sim \mathcal{N}_q(\mathbf{0},\bSigma).
\end{equation}
We assume that the response variables can be split into two different groups $\bY_1$ and $\bY_2$ of dimensions $n\times \qg$ and $n\times (q-\qg)$, respectively, where $\qg \in \{ 2,\ldots,q-1 \}$.
Moreover, we assume that the relationship between $\bY_1$ and $\bX$ admits a low-rank structure, while the regression of $\bY_2$ on $\bX$ has full-rank. Under this assumption, the coefficient matrix $\bC \in\R^{p\times q}$ can be partitioned as $\bC = [\bC_1, \bC_2]$, with $\bC_1 \in \R^{p\times\qg}$ having reduced rank $r = \rank(\bC_1) \leq \min(p,\qg)-1$ and $\bC_2\in\R^{p\times(q-\qg)}$ with full rank $r_2 = \rank(\bC_2) = \min(p,q-\qg)$.

Therefore, the model in Eq.~\eqref{eq:linearmodel} can be represented with partitioned matrices as:
\begin{equation}
\label{eq:rrmodel}
	\left[\bY_1,\, \bY_2\right] = \bX\left[\bC_1,\, \bC_2\right] + \left[\bE_1,\, \bE_2\right].
\end{equation}
Notice that each of the $n$ response vectors $\by_{(i)}$, $i=1,\ldots,n$, is of the form $\by_{(i)} = ({y_{i,1}, \ldots, y_{i,\qg}}, {y_{i,\qg+1}, \ldots,y_{i,q}})' \in \R^q$.
We write $\be_{(i)} = (\be_{1i}',\be_{2i}')$ with $\be_{1i} = (e_{i,1}, \ldots, e_{i,\qg})'$ and $\be_{2i} =(e_{i,\qg+1}, \ldots, e_{i,q})'$, and assume $\be_{(i)}$ is normally distributed with mean zero and a partitioned covariance matrix $\bSigma=\cov(\be_{(i)})$ such that $\bSigma_{11}=\cov(\be_{1i})$, $\bSigma_{22}=\cov(\be_{2i})$, and $\bSigma_{12}= \cov(\be_{1i},\be_{2i})$.

\subsection{Prior specifications}
\label{sec:priors}

The model in Eq.~\eqref{eq:rrmodel} classifies the response variable into two groups. Differently from \cite{reinsel2006partially}, we assume the grouping structure to be unknown and aim at inferring it from the data.
Therefore, we introduce a binary vector $\bgamma \in \{0,1\}^q$ to categorise the responses into the low-rank and the full-rank groups. As we lack any prior information regarding the criteria for this classification, we assume that each element $\gamma_j$, $j=1,\ldots,q$, follows independently a Bernoulli prior distribution with probability $\rho$ of being assigned to the low-rank group. Consequently, the joint prior distribution on $\bgamma$ is:
\begin{equation}
\label{eq:priorgamma}
    p(\bgamma|\rho) = \left[ \prod_{j=1}^q \text{Bern}(\gamma_j|\rho)\right] \I\!\left( 1\!<\!q_{\bgamma}\!<\!q \right),
\end{equation}
where %
$q_{\bgamma}=\sum_{j=1}^{q} \gamma_j$, and $\rho \in (0,1)$ is the prior probability of being assigned to the low-rank group. The constraint imposed by the indicator function in Eq.~\eqref{eq:priorgamma} allows for the existence of the low-rank group and, thus, of a PRR model.
In fact, if $\qg = 1$, $\bY_1$ comprises only one response, making it a full-rank submatrix. Conversely, when $\qg = q$, all responses are part of the low-rank group, which collapses into the standard RR model.
Additionally, we employ a hierarchical prior structure, where $\rho$ is assigned a Beta prior distribution, $\rho \sim \mathcal{B}e(\rho | \underline{a}_\rho,\underline{b}_\rho)$.

Moving to the coefficients matrix, $\bC$, we can provide a specification for both the low and the full rank matrix. In particular,
the submatrix of coefficients $\bC_1$ is assumed to have reduced rank $r \leq \rmax = \min(p,\qg)-1$, which depends on the binary parameter $\bm{\gamma}$. Therefore, conditional on $\qg$ (hence on $\bgamma$), we assume an uninformative uniform prior distribution for $r$ over the discrete set $\{ 1, \ldots, \rmax\}$, that is $r | \bgamma \sim \mathcal{U}(r | \{1, \ldots, \rmax\})$.

Given that $\bC_1$ is a low-rank matrix, we can express it as the product of two full-rank matrices $\bA\in \R^{\qg\times r}$ and $\bB\in \R^{p\times r}$, such that $\bC_1=\bB\bA'$. This decomposition is not unique, since for any orthogonal $r \times r$ matrix $\mathbf{P}$, we have that $\bC_1 = (\bB\mathbf{P})(\mathbf{P}'\bA')$. To achieve a unique decomposition of $\bC_1$, we follow \cite{geweke1996bayesian} and impose an identifying restriction by assuming the first $r$ rows of $\bA$ are the identity matrix $\bI_r$, that is
\begin{equation}
\label{eq:A}
    \bA = \begin{bmatrix} \bI_r \\  \bF \end{bmatrix},
\end{equation}
where $\bF$ is a $(\qg-r) \times r$. Denoting with $\vec(\cdot)$ the vectorisation operator, we assume a multivariate Gaussian prior distribution on $\balpha_F = \vec(\bF')$, that is
\begin{equation}
    \balpha_F | \bgamma,r \sim \mathcal{N}_{(\qg-r)r} (\balpha_F | \bzero,\Lambdau_{\balpha}),
\end{equation}
where $\Lambdau_{\balpha} = \underline{a} \, \bI_{(\qg-r)r}$, for fixed $\underline{a} > 0$.
Similarly, defining $\bbeta = \vec(\bB)$ and $\bdelta = \vec(\bC_2)$, we assume the multivariate Gaussian prior distributions:
\begin{align}
    \bbeta | \bgamma,r & \sim \mathcal{N}_{pr}(\bbeta | \bzero, \Lambdau_{\bbeta}), \\
    \bdelta | \bgamma & \sim \mathcal{N}_{p(q-\qg)}(\bdelta | \bzero, \Lambdau_{\bdelta}),
\end{align}
where $\Lambdau_{\bbeta} = \underline{b}\, \bI_{pr}$, and $\Lambdau_{\bdelta} = \underline{d} \, \bI_{p(q-\qg)}$, for fixed $\underline{b}, \, \underline{d} > 0$.
Lastly, we adhere to the conventional practice and assign a conjugate inverse Wishart prior to $\bSigma$, that is $\bSigma \sim \mathcal{IW}_q(\bSigma | \nuu,\Psiu)$, with $\nuu$ and $\Psiu$ being the fixed degrees of freedom and scale matrix.

%%%%%%%%%%%%%%%%%%%%%%%%%%%%%%%%%%%%%%%%%%%%%%%%%%%%%%%%%%%%%%%%%%%%%%%%%%

\section{Posterior sampling} 
\label{sec:posteriors}

In this section, we design an MCMC algorithm to draw samples from the joint posterior distribution $p(\bA,\bB,\bC_2,\bSigma,r,\bgamma,\rho|\bY)$.
The most critical issue to tackle is that the dimensions of the matrices $\bA$, $\bB$, and $\bC_2$ depend on the states of $\bgamma$ and $r$, which implies that the dimension of the parameter space may change across the iterations of the MCMC algorithm.
Consequently, the traditional Gibbs sampler is invalid in this setting, whereas a reversible jump MCMC, although theoretically feasible, poses significant challenges in terms of implementation and proper execution \citep{robert1999monte}.
To address this challenge, we implement a partially collapsed Gibbs sampler \citep[PCG, see][]{vandyk2008partially}, which generalises the composition of the conditional distributions in Gibbs samplers, relying on three basic tools: marginalization, permutation, and trimming.
Specifically, we avoid the need for transdimensional samplers by drawing $(\bgamma,r)$ from a joint distribution marginalised over the parameters $(\bA,\bB,\bC_2)$ whose size depends on $(\bgamma,r)$. Subsequently, we sample $(\bA,\bB,\bC_2)$ conditionally on the updated values of $(\bgamma,r)$.
The entire sampling process is summarised in Algorithm~\ref{alg:PCG}.

In the remainder of this section, we describe the procedures adopted to integrate out $(\bA,\bB,\bC_2)$ from the likelihood, then we explain the main computational challenges and the proposed solutions.
The first problem arises in Step 1, the most computational intensive step, where sampling $\bgamma$ entails the exploration of a $2^q$ dimensional parameter space and the computation of analytically intractable integrals, a limitation also encountered in Step 2. The second issue concerns the dimensions of matrices $\bA$, $\bB$, and $\bC_2$ in practically implementing Steps 3 and 4, since their size depends on the states of $(\bgamma,r)$, producing a potential incompatibility between the dimension of the parameters at the previous MCMC iteration and the current one corresponding to the updated $(\bgamma,r)$.

\begin{algorithm} 
\caption{PCG for Bayesian PRR model}
\label{alg:PCG}
\begin{algorithmic}[1]
\STATE Sample $\bgamma$ from $p(\bgamma|\bY,\bSigma,\rho)$.
\STATE Sample $r$ from $p(r|\bgamma,\bY,\bSigma)$.
\STATE Sample $\bdelta = \vec(\bC_2)$ from $p(\bdelta|\bY,\bSigma,\bgamma,\bA,\bB) = \mathcal{N}_{p(q-\qg)}(\overline{\bmu}_\delta, \overline{\bSigma}_\delta)$.
\STATE Sample $\balpha_\bF = \vec(\bF')$ from $p(\balpha_F|\bY,\bSigma,\bgamma,r,\bB,\bC_2) = \mathcal{N}_{(\qg-r)r}(\overline{\bmu}_\alpha, \overline{\bSigma}_\alpha)$,\\ then set $\bA = \left[\bI_r, \bF'\right]'$.
\STATE Sample $\bbeta = \vec(\bB)$ from $p(\bbeta|\bY,\bSigma,\bgamma,r,\bA,\bC_2) = \mathcal{N}_{pr}(\overline{\bmu}_\beta, \overline{\bSigma}_\beta)$.
\STATE Sample $\bSigma$ from $p(\bSigma|\bY,\bgamma,\bA,\bB,\bC_2) = \mathcal{IW}_q(\nuo, \Psio)$.
\STATE Sample $\rho$ from $p(\rho | \bgamma) = \mathcal{B}e(\overline{a}_\rho, \overline{b}_\rho)$.
\end{algorithmic}
\end{algorithm}

\subsection{Preliminaries}

Let us define the likelihood function as:
\begin{equation} 
\label{eq:likelihood}
    f(\bY|\bA,\bB,\bC_2,\bSigma,r,\bgamma) = \frac{1}{(2\pi)^{nq/2}\abs{\bSigma}^{n/2}}\exp\Bigg\{ -\frac{1}{2}\norm{(\bY-\bX\bC)\bSigma^{-1/2}}_F^2 \Bigg\},
\end{equation}
where $\bC=[\bC_1,\bC_2]$, $\bC_1=\bB\bA'$, and $\norm{\cdot}_F$ denotes the Frobenius norm.
The first step involves explicitly expressing the likelihood in terms of $\bC_1$ and $\bC_2$. Therefore, we exploit the partitioning of $\bC$ and rewrite the model in Eq.~\eqref{eq:rrmodel} equivalently as
\begin{equation}
\label{eq:rrmodel2}
	\bY = \bX \bC_1 \bV_1 + \bX \bC_2 \bV_2 + \bE,
\end{equation}
where $\bV_1=\left[\bI_{q_\gamma}, \bzero_{q_\gamma \times (q-\qg)}\right]$, and $\bV_2=\left[\bzero_{(q-\qg) \times q_\gamma},\bI_{q-\qg}\right]$.
By vectorising Eq.~\eqref{eq:rrmodel2}, we obtain
\begin{equation} 
\label{eq:vecii}
    \by = \bU_1 \bc_1 + \bU_2 \bdelta + \be,
\end{equation}
where $\by = \vec(\bY)$, $\bc_1 = \vec(\bC_1)$, $\bdelta = \vec(\bC_2)$, $\be=\vec(\bE)$, and $\bU_i = \bV_i' \otimes \bX$, for each $i=1,2$. % and $\bU_2 = \bV_2' \otimes \bX$.
It follows that $\by | \bA,\bB,\bC_2,\bSigma,r,\bgamma \sim \mathcal{N}_{nq}(\by | \bU_1 \bc_1 + \bU_2 \bdelta, \tilde{\bSigma})$, where $\tilde{\bSigma} = \bSigma \otimes \bI_n$. 

Under the vectorised model in Eq.~\eqref{eq:vecii}, the likelihood can be marginalized over $\bC_2$ analytically to obtain
\begin{equation}
\label{eq:marglike}
\begin{split}
    f(\by | \bA,\bB,\bSigma, \bgamma,r) & = \int f(\bY|\bA,\bB,\bC_2,\bSigma,\bgamma,r)\, p(\bC_2|\bgamma) \, \d\bC_2 \\
     & = \int \mathcal{N}_{nq}(\by | \bU_1 \bc_1 + \bU_2 \bdelta, \tilde{\bSigma}) \, \mathcal{N}_{p(q-\qg)}(\bdelta | \mathbf{0}, \Lambdau_{\bdelta}) \, \d \bdelta \\
     & = \mathcal{N}_{nq}(\by | \bU_1 \bc_1, \, \tilde{\bSigma} + \bU_2 \Lambdau_{\bdelta} \bU_2').
\end{split}
\end{equation}
This distribution represents the starting point in the design of the proposed PCG sampler.

\subsection{Sampling the response allocation, \texorpdfstring{$\bgamma$}{gamma}, and rank, $r$}
\label{sec:gamma}

Starting from Eq.~\eqref{eq:marglike}, we are left with the task of marginalising $\bA$ and $\bB$. As analytical integration is unfeasible, we obtain an approximation to the (marginal) posterior of $\bgamma$ via the Laplace method, which provides a trade-off between computational speed and accuracy \citep[e.g., see][among others]{tierney1986accurate,tierney1989fully,kass1995bayes}. Then, a sample from the approximate posterior is obtained through the Metropolized Shotgun Stochastic Search (MSSS) algorithm \citep{hans2007shotgun}. Our strategy is similar in spirit to \cite{yang2022rrr}, which is concerned with rank estimation in a RR model.

The posterior of $\bgamma$ given by Bayes' theorem after integrating out $\bA$,\,$\bB$, $\bC_2$ and $r$ is
\begin{equation}
\label{eq:posteriorgamma}
    p(\bgamma|\bY,\bSigma,\rho) = \frac{f_\gamma(\bY|\bSigma,\bgamma) p(\bgamma|\rho)}{\sum_{\bgamma^\dag\in\{0,1\}^q} f_\gamma(\bY|\bSigma,\bgamma^{\dag}) p(\bgamma^{\dag}|\rho)},
\end{equation}
where $f_\gamma(\bY|\bSigma,\bgamma)$ is obtained from Eq.~\eqref{eq:likelihoodgamma} by marginalising over $\bA$, $\bB$, and $r$, that is
\begin{align}
\label{eq:likelihoodgamma}
    f_\gamma (\bY|\bSigma,\bgamma) & = \sum_{r=1}^{\rmax}\frac{1}{\rmax} \iint f(\bY|\bA,\bB,\bSigma,\bgamma,r) \, p(\bA,\bB|r,\bgamma) \, \d \bA \, \d \bB \\
    \notag
    & = \sum_{r=1}^{\rmax}\frac{1}{\rmax} f_r(\bY | \bSigma, \bgamma,r).
\end{align}
Note that the integration with respect to $r$ is performed analytically since the latter is a discrete parameter with finite support.
Conversely, we use the Laplace method \citep{raftery1995bayesian} to approximate the analytically intractable integration of $\bA,\bB$ to obtain 
\begin{equation}
\label{eq:loglikelihoodgammar}
    \log f_r(\bY|\bSigma,\bgamma,r) \approx \log f(\bY|\hat{\bA},\hat{\bB},\bSigma,\bgamma,r) -\frac{1}{2}(pr+(\qg-r)r)\log n,
\end{equation}
where $\hat{\bA}$ and $\hat{\bB}$ are the maximum likelihood estimators (MLEs) of $\bA$ and $\bB$, given $r$ and $\bgamma$. The main difficulty relies on the computation of these MLEs, although \cite{reinsel2022book} make available these estimators for i.i.d. Gaussian errors, our setting differs from this baseline in two main aspects.
First, the Gaussian density in Eq.~\eqref{eq:marglike} has a covariance matrix $\bSigma_\by = \tilde{\bSigma} + \bU_2 \Lambdau_{\bdelta} \bU_2'$ that incorporates heteroscedastic errors through the dependency of $\bU_2$ on $\bV_2$. As a consequence, the block of the covariance matrix corresponding to $\bC_2$ introduces a different variance for each observation.
Second, we are imposing an identification restriction on the matrix $\bA$ and an additional restriction on the vectorised linear model's coefficient via the binary matrix $\bV_1$. 

The first restriction requires the first $r$ rows of $\bA$ to be the identity matrix. The second restriction pertains to the representation of $\bY$ in Eq.~\eqref{eq:rrmodel2} as the sum of a low-rank component and its full-rank counterpart, which clearly demands the introduction of zero factors through $\bV_1$ and $\bV_2$ to accommodate the desired structure.
Consequently, an alternative procedure to compute the MLEs is required.

\cite{hansen2002generalized} proposed an ML estimation technique for a general class of reduced rank regression models (called GRRR), including models with a generic structure of the covariance matrix and potential restrictions on the coefficient matrices.
We exploit the GRRR setting to accommodate the heteroscedasticity and the aforementioned restrictions in the computation of the MLEs.

The GRRR problem considers the regression given by the vectorised model of Eq.~\eqref{eq:marglike}
\begin{equation}
    \by_{(i)} = \bV_1'\bA\bB' \bx_{(i)} + \tilde{\be}_{(i)},
\end{equation}
where $\tilde{\be}_{(i)}$ is the $i$th column of $\tilde{\bE} \in \mathbb{R}^{q\times n}$ and $\vec(\tilde{\bE}) \sim \mathcal{N}_{nq}(\bzero,\bSigma_\by)$, 
% where $\tilde{\be}_{(i)}'$ is the $i$th row of $\bE' \in \mathbb{R}^{n\times q}$ and $\vec(\tilde{\bE}) \sim \mathcal{N}_{nq}(\bzero,\bSigma_\by)$,
subject to the restriction
\begin{equation}
    \vec(\bV_1'\bA) = \bG\psi + \mathbf{g},
\end{equation}
where $\psi$ is the true vector of parameters to be estimated, 
$G$ is the binary $qr \times r(\qg-r)$ matrix $\bG$ and $\mathbf{g}$ is the $qr$-dimensional binary vector of restrictions (see Section 2 of the Supplement for details). 
Then, the MLEs following the GRRR method are obtained as
\begin{align}
\label{eq:mlesa}
    \hat{\balpha}_{\bV_1} &= \vec(\bV_1'\bA) = \bG (\bG'\bM_\bB \bG)^{-1} \bG'(\mathbf{n}_\bB-\bM_\bB \mathbf{g}) + \mathbf{g},\\
    %\hat{\balpha}_{V_1} &= \left[ (XB \otimes I_q)' \Sigma_{\by}^{-1} (XB \otimes I_q) \right]^{-1} (XB \otimes I_q)' \Sigma_{\by}^{-1} \vec(\bY'), \\
\label{eq:mlesb}
    \hat{\bbeta} &= \vec(\bB)  = \bM_\bA^{-1} \mathbf{n}_\bA,
\end{align}
where $\bM_\bB=(\bX\bB \otimes \bI_q)' \tilde{\bSigma}_\by^{-1} (\bX\bB \otimes \bI_q)$,
$\mathbf{n}_\bB = (\bX\bB \otimes \bI_q)' \tilde{\bSigma}_{\by}^{-1} \vec(\bY')$,
$\bM_\bA = \mathbf{K}_{p,r}' (\bX \otimes \bV_1'\bA)' \tilde{\bSigma}_\by^{-1} (\bX \otimes \bV_1'\bA) \mathbf{K}_{p,r}$,
$\mathbf{n}_\bA = \mathbf{K}_{p,r}' (\bX \otimes \bV_1'\bA)' \tilde{\bSigma}_{\by}^{-1} \vec(\bY')$,
$\tilde{\bSigma}_{\by} = \mathbf{K}_{n,q} \bSigma_{\by} \mathbf{K}_{n,q}'$,
and $\mathbf{K}_{m,n}$ is the $mn \times mn$ commutation matrix, which transforms the vectorisation of a matrix $\bM \in \mathbb{R}^{m\times n}$ into the vectorisation of its transpose, such that $\mathbf{K}_{m,n} \vec(\bM) = \vec(\bM')$.

Noticing that the expressions in Eq. \eqref{eq:mlesa} and \eqref{eq:mlesb} depend on each other, the practical implementation of the GRRR method is done in a recursive algorithm. After a random initialization of the parameter values, $\balpha_{\bV_1}$ and $\bbeta$ are iteratively updated until convergence.
Once a solution $\hat{\balpha}_{\bV_1},\hat{\bbeta}$ is obtained, it suffices to transform the vectorised MLEs back to their matrix forms $\bV_1 \hat{\bA}'$ and $\hat{\bB}$ to obtain the MLEs of the low-rank coefficient matrix as $\hat{\bC}_1 = \hat{\bB} (\bV_1')^+ \bV_1' \hat{\bA} = \hat{\bB} \hat{\bA}'$, where $\bM^+$ refers to the Moore-Penrose pseudoinverse of $\bM$.\footnote{$\bV_1$ is not an invertible matrix given that its dimensionality is $\qg \times q$, with $\qg < q$.}

Consequently, we obtain the approximation
\begin{equation}
    f_\gamma(\bY|\bSigma,\bgamma) \approx \sum_{r=1}^{\rmax}\frac{1}{\rmax} \tilde{f}_r(\bY|\bSigma,\bgamma,r) \equiv \tilde{f}_\gamma(\bY|\bSigma,\bgamma),
\end{equation}
where $\tilde{f}_r(\bY|\bSigma,\bgamma,r)$ is the Laplace approximation of $f_r(\bY|\bSigma,\bgamma,r)$ obtained from Eq.~\eqref{eq:loglikelihoodgammar} to the integral in Eq.~\eqref{eq:likelihoodgamma}.
Therefore, the  posterior distribution of $\bgamma$ is approximated by
\begin{equation}
\label{eq:posteriorgammaapprox}
    \tilde{p}(\bgamma|\bY,\bSigma,\rho) = \frac{\tilde{f}_\gamma(\bY|\bSigma,\bgamma) p(\bgamma|\rho)}{\sum_{\bgamma^\dag\in\{0,1\}^q} \tilde{f}_\gamma(\bY|\bSigma,\bgamma^{\dag}) p(\bgamma^{\dag}|\rho)}.
\end{equation}

Given that, $\bgamma$ is a $q$-dimensional binary vector, the collection of all possible configurations for the response allocation encompasses $2^q$ distinct elements, which quickly becomes exceedingly large even for moderate $q$.
This calls for the adoption of an approximate method to sample $\bgamma$. Given the discreteness of the support of $\bgamma$, we use a Metropolized Shotgun Stochastic Search (MSSS) algorithm proposed by \cite{hans2007shotgun}.
The MSSS approach explores regions of the high-dimensional parameter space by examining a selection of neighbours of the current iteration's $\bgamma$ and rapidly identifying those with high posterior probability. Defining the set of neighbours to contain only a subset of all the possible values of $\bgamma$ allows for a trade-off between the space exploration speed and the computational time.
Similar to \cite{yang2022rrr}, we take the neighbourhood to be every binary vector that is a one-variable change to the current allocation $\bgamma$ and at the same time complies with the existence of a low-rank group.
For example, if $\bgamma = (1,0,1,0)$, then $\nbd(\bgamma) = \left\{(1,1,1,0), (1,0,1,1) \right\}$, but $(0,1,1,0)$, $(0,0,1,0)$ and $(1,0,0,0)$ are not a neighbours.\footnote{The element $(0,1,1,0)$ is not a neighbour because two variables changed; $(0,0,1,0)$ and $(1,0,0,0)$ are not neighbours because they violate the constraint $\qg \in \{1,\ldots,q-1\}$.}
This restriction improves computational efficiency compared to an unrestricted neighbourhood comprising all elements while allowing the (reduced) exploration of the space.
We define a proposal distribution by
\begin{equation}
\label{eq:ggamma}
g(\bgamma|\bgamma^{(m)}) \propto \tilde{p}(\bgamma|\bY,\bSigma,\rho) \, \I \left(\bgamma \in \nbd(\bgamma^{(m)})\right),
\end{equation}
where $\bgamma^{(m)}$ is the value of $\bgamma$ at the $m$th iteration of the MCMC.
Summarising, the first step of the proposed PCG sampler generates a draw from the marginal posterior $p(\bgamma | \bY,\bSigma,\rho)$ with the following procedure:
\begin{enumerate}[label=\arabic*.]
    \item Generate $\bgamma^*$ from $g(\bgamma|\bgamma^{(m)})$.
    \item Accept $\bgamma^{(m+1)}=\bgamma^*$ with probability 
        \begin{equation}
        \label{eq:probgamma}
            \rho_\gamma = \min\left\{1,
            \frac{\sum_{\bgamma\in\nbd(\bgamma^{(m)})} \tilde{f}_\gamma(\bY|\bSigma,\bgamma) p(\bgamma|\rho)}
            {\sum_{\bgamma^{\dag}\in\nbd(\bgamma^*)} \tilde{f}_\gamma(\bY|\bSigma,\bgamma^{\dag}) p(\bgamma^{\dag}|\rho)}
            \right\},
        \end{equation}
        and otherwise, set $\bgamma^{(m+1)}=\bgamma^{(m)}$.
\end{enumerate}

Similarly to $\bgamma$, the conditional posterior of $r$ is approximated through the Laplace method. Specifically, we compute the approximated posterior
\begin{equation}
\label{eq:posteriorrapprox}
    \tilde{p}(r|\bY,\bSigma,\bgamma) = \frac{\tilde{f}_r(\bY|\bSigma,\bgamma,r)p(r|\bgamma)}{\sum_{r^{\dag}=1}^{\rmax} \tilde{f}_r(\bY|\bSigma,\bgamma, r^{\dag}) p(r^{\dag}|\bgamma)},
\end{equation}
where $\tilde{f}_r(\bY|\bSigma,\bgamma,r)p(r|\bgamma)$ is the same as in Eq.~\eqref{eq:loglikelihoodgammar}.
Then, a new value of $r$ is sampled from the discrete distribution on $\{ 1,\ldots,\rmax\}$ with the probabilities given in Eq.~\eqref{eq:posteriorrapprox}.

\subsection{Sampling the matrices \texorpdfstring{$\bA$}{a} and \texorpdfstring{$\bC_2$}{a}}
\label{sec:diminc}

The proposed PCG sampler introduces a particular challenge related to the dimensions of matrices $\bA$, $\bB$, and $\bC_2$ in Steps 3 and 4 of Algorithm~\ref{alg:PCG}. Suppose that at the end of iteration $m$, we have generated $(\bgamma^{(m)}, r^{(m)}, \bA^{(m)}, \bB^{(m)}, \bC_2^{(m)})$. Then, at iteration $m+1$, we obtain new values $(\bgamma^{(m+1)}, r^{(m+1)})$ in Steps 1 and 2.
Afterwards, we shall update $\bC_2$ by sampling $\bdelta^{(m+1)}$ conditioned on the value $\bgamma^{(m+1)}$ just generated. To this aim, notice that $\bC_2^{(m+1)}$ should be of ``new'' dimension $p \times (q-q_{\bgamma^{(m+1)}})$, and its posterior distribution depends on the matrix $\bC_1$, which should have ``new'' dimension $p \times q_{\bgamma^{(m+1)}}$. However, the matrix $\bC_1^{(m)}$ available at this step is of dimension $p \times q_{\bgamma^{(m)}}$.
An analogous issue is encountered in Step 4 when sampling $\bA^{(m+1)}$, as the posterior of the latter parameter would require a matrix $\bB$ with ``new'' dimension $p \times r^{(m+1)}$, whereas the available matrix $\bB^{(m)}$ has $r^{(m)}$ columns.

It is worth emphasising that these dimensionality issues stem from considering $(\bgamma,r)$ as parameters to be estimated, thus varying quantities across the MCMC iterations. Moreover, changing the order of the Gibbs steps (while keeping the PCG sampler) would not circumvent the problem.

A possible way out of the issue in Step 3 can be found by recalling the decomposition of the coefficient matrix in Eq.~\eqref{eq:rrmodel}, that is $\bC^{(m)} = \big[ \bC_1^{(m)}, \bC_2^{(m)} \big]$. Importantly, for any iteration $m=1,\ldots,M$, this matrix of coefficients has a fixed dimension $p \times q$, whereas the number of columns of $\bC_1^{(m)}$ and $\bC_2^{(m)}$ are possibly changing across iterations in consequence of varying $\bgamma^{(m)}$. Hence, to sample $\bdelta^{(m+1)}$ conditioned on $\bgamma^{(m+1)},\bA^{(m)},\bB^{(m)}$, we construct an auxiliary matrix $\bC_{1*}^{(m)}$ that is consistent with the newly sampled value of $\bgamma^{(m+1)}$, formed by selecting the first $q_{\bgamma^{(m+1)}}$ columns of the available $\bC^{(m)}$:
\begin{equation}
    \bC_{1*}^{(m)} = \left[ \bC_{\bullet 1}^{(m)}, \ldots, \bC_{\bullet q_{\bgamma^{(m+1)}}}^{(m)} \right],
\label{eq:c1star}
\end{equation}
where $\bC_{\bullet j}^{(m)}$ is the $j$th column of matrix $\bC^{(m)}$. At iteration $m+1$, the auxiliary matrix $\bC_{1*}^{(m)}$ is used to compute the updated parameters of the posterior distribution of $\bdelta^{(m+1)}$.

To address the dimensionality inconsistency in Step 4, let us recall the restriction in Eq.~\eqref{eq:A}; then, substituting $\bA^{(m)} = \left[\bI_{r^{(m)}},\bF^{(m)\prime}\right]'$ and $\bB^{(m)}$ in the low-rank matrix representation yields
\begin{equation}
    \bC_1^{(m)} = \left[ \bB^{(m)}, \bB^{(m)}\bF^{(m)\prime}\right].
\label{eq:c1}
\end{equation}
Based on Eq.~\eqref{eq:c1}, it is evident that the first $r^{(m)}$ columns of $\bC_1^{(m)}$ coincide with the matrix $\bB^{(m)}$. Consequently, to update $\bA^{(m+1)}$, we define the auxiliary matrix $\bB_{*}^{(m)}$ as:
\begin{equation}
    \bB_{*}^{(m)} = \left[ \bC_{\bullet 1}^{(m)}, \ldots, \bC_{\bullet r^{(m+1)}}^{(m)} \right].
\label{eq:bstar}
\end{equation}
It is important to remark that the auxiliary matrices $\bC_{1*}^{(m)}$ and $\bB_{*}^{(m)}$ have appropriate dimensions, corresponding to the updated values $\bgamma^{(m+1)}$ and $r^{(m+1)}$, and contain elements already available at iteration $m+1$, that is $\bC^{(m)}$, $\bB^{(m)},\bF^{(m)}$. %Overall, this enables us to sample the new values $\bC_2^{(m+1)}$ and $\bA^{(m+1)}$.

In more detail, in Step 3, the full-rank coefficient matrix is sampled in vectorised form. Denoting $\bc_{1*} = \vec(\bC_{1*})$, the posterior distribution of $\bdelta$ is proportional to the multivariate Gaussian distribution $p(\bdelta | \bY,\bSigma,\bgamma, \bc_{1*}) \propto p(\bdelta) \, p(\by | \bSigma,\bdelta, \bc_{1*}) \sim \mathcal{N}_{p(q-\qg)}(\bdelta | \bmuo_{\bdelta}, \Lambdao_\delta)$ with mean $\bmuo_{\bdelta} = \Lambdao_{\bdelta} \bU_2' \tilde{\bSigma}^{-1} (\by - \bU_1 \bc_{1*})$ and covariance matrix $\Lambdao_{\bdelta} = (\Lambdau_{\bdelta}^{-1} + \bU_2' \tilde{\bSigma}^{-1} \bU_2)^{-1}$.

The update of $\bA = \left[\bI_r,\bF'\right]'$ given $(\bY, \bgamma, r, \bSigma, \bB, \bC_2)$ is performed by sampling $\balpha_\bF = \vec(\bF')$. The posterior distribution of $\balpha_\bF$ is proportional to the multivariate Gaussian distribution $p(\balpha_\bF | \bY,\bSigma,\bgamma,r,\bC_2, \bB_{*}) \propto p(\balpha_\bF|\bgamma,r) \, p(\bY|\bSigma,\bA,\bB_{*},\bC_2) \sim \mathcal{N}_{(\qg-r)r}(\balpha_\bF | \bmuo_{\balpha}, \Lambdao_{\balpha})$ with mean $\bmuo_{\balpha} = \Lambdao_{\balpha}\big( \mathbf{m}_J - \bG_{\left[J,J\right]} \mathbf{v} \big)$ and covariance matrix $\Lambdao_{\balpha} = \big(\Lambdau_{\balpha}^{-1} + \bG_{\left[J,J\right]} \big)^{-1}$, where 
$\mathbf{v} = \vec(\bI_r)$, 
$\mathbf{m} = \bM_{\balpha}' \tilde{\bSigma}^{-1} \tilde{\by}_2$, 
$\bG = \bM_{\balpha}' \tilde{\bSigma}^{-1} \bM_{\balpha}$, 
$\tilde{\by}_2 = \by - \bU_2 \bdelta$, 
and 
$\bM_{\balpha} = \bU_1 (\bI_{\qg} \otimes \bB)$.
Moreover, $\bG_{\left[J,J\right]}$ indicates the $J$th row and the $J$th column in $\bG$ for the sequence $J = \left\{r^2+1,r^2+2,\ldots,\qg r\right\}$.

\subsection{Sampling the other parameters}
\label{sec:otherparam}

Regarding Steps (5)--(7) of Algorithm~\ref{alg:PCG}, the conditional posterior distribution of $\bbeta$, given $(\bY,\bgamma,r,\bSigma,\bA,\bC_2)$, is proportional to the multivariate Gaussian distribution $\mathcal{N}_{pr}(\bbeta | \bmuo_{\bbeta}, \Lambdao_{\bbeta})$, where $\Lambdao_{\bbeta} = (\Lambdau_{\bbeta}^{-1} + \bM_{\bbeta}' \tilde{\bSigma}^{-1} \bM_{\bbeta})^{-1}$ and $\bmuo_{\bbeta} = \Lambdao_{\bbeta} \bM_{\bbeta}' \tilde{\bSigma}^{-1} \tilde{\by}_2$, with $\bM_{\bbeta} = \bU_1 (\bA \otimes \bI_p)$.

The conditional posterior of the innovation covariance matrix $\bSigma$, given $(\bA, \bB, \bC_2, \bY)$, is the inverse Wishart $\mathcal{IW}_q(\bSigma|\nuo,\Psio)$, where $\nuo = \nuu + n$ and $\Psio = \Psiu + (\bY - \bX \bC)'(\bY - \bX \bC)$.
Finally, the posterior distribution of $\rho$, the probability of a response variable belonging to the low-rank group, is the Beta distribution $\mathcal{B}e(\rho | \overline{a}_\rho, \overline{b}_\rho)$, with $\overline{a}_\rho = \underline{a}_\rho + \qg$, and $\overline{b}_\rho = \underline{b}_\rho + q - \qg$.
A detailed derivation of the posterior distributions is available in the Supplement. %(Section 1).

%%%%%%%%%%%%%%%%%%%%%%%%%%%%%%%%%%%%%%%%%%%%%%%%%%%%%%%%%%%%%%%%%%%%%%%%%%

\section{Simulation study}
\label{sec:sim}

This section is devoted to examining the proposed model's performance in terms of the ability to recover the true group allocation, that is, the classification of variables into the low- and full-rank groups in different simulation studies. Next, the performance of the sampler is tested about rank estimation and the recovery of the overall coefficient matrix, $\bC$.

The data was generated from the multivariate linear model $\bY_0 = \bX\bC_0 + \bE_0$. The rows of $\bX$ were independently drawn from $\mathcal{N}(\bzero,\bI_p)$ and the rows of $\bE_0$ were drawn from $\mathcal{N}(\bzero,\bSigma_0)$, where the covariance matrix $\bSigma_0$ is diagonal with elements sampled from $\mathcal{U}(0.5,1.75)$. We work with centred responses and exclude the intercept term for simplicity.
To generate the coefficient matrix $\bC_0$, we first recall its partition into the low-rank and full-rank matrices $\bC_1 = \bB\bA'$ and $\bC_2$, then draw each free entry of $\bA$, $\bB$, and $\bC_2$ from a standard Gaussian. 
Notice that the dimensions of these matrices depend on the fixed number of responses in the low-rank group, $\qg < q$, and the true rank, $r \leq \rmax$.

The allocation of the response variables to the reduced-rank group is randomly selected given $\qg$, and represented by the binary vector $\bgamma$. The columns of the matrix $\bY_0$ are then permuted following the allocation imposed by $\bgamma$.
The response matrix so generated, $\bY$, need not necessarily be partitioned as $\bY_0 = [\bY_1,\bY_2]$, which is the representation postulated by our BPRR model.
For example, if $q=5$ and $\qg=3$, the data generating process (DGP) initially comprises the partitioned response matrix $\bY_0 = [\bY_1,\bY_2]$, with $\bY_1=[\by_1, \by_2, \by_3] \in \R^{n\times 3}$ and $\bY_2 = [\by_4, \by_5] \in \R^{n\times 2}$. However, after a random $\bgamma$ has been generated, say $\bgamma_0 = (0,1,0,1,1)$, then the final generated response matrix that is fed into our model is $\bY = [\by_4,\by_1,\by_5,\by_2,\by_3]$, aiming to reorder the responses to their true form, $\bY_0$, if $\hat{\bgamma}$ is estimated correctly.  

The hyperparameters are set to consider noninformative priors, specifically   $\underline{a}_\rho = \underline{b}_\rho = 1$, $\underline{a} = \underline{b} = \underline{d} = 0.5$, $\nuu = q+1$, and $\Psiu = \bI_q$. The starting value of $\bSigma$ is the identity matrix, while the coefficient matrix and the response allocation vector are initialised at random.

We consider different simulation settings with varying dimensionality, number of low-rank responses and true rank. Our method (BPRR) is compared with the following competitors: full-rank (FR), \textit{full} low-rank (RR), and \textit{pre-specified allocation} partial low-rank (PRR*). 
The first one is a standard linear regression model, where no low-rank structure is assumed. 
The full low-rank concerns a usual reduced-rank regression without an imposed partition. 
The last competitor is a partially reduced-rank regression model in which the low-rank group is fixed at random, a feature that serves two purposes. 
First, it accommodates scenarios where the researcher might have prior knowledge about the grouping structure, enabling the estimation procedure of the partial reduced-rank model to be conducted with a constraint by the imposed $\bgamma$. 
Second, it allows us to examine whether the automatic model selection in BPRR offers advantages over a random grouping choice.

The performance of the estimator $\hat{\bC}$ of the coefficient matrix is evaluated using the mean squared error, \mbox{$\text{MSE}=\|\hat{\bC}-\bC_0\|_\text{F}^2/(pq)$}, where $\hat{\bC}$ is the posterior mean of the predicted coefficients in their original ordering, warranting a fair comparison if the estimated allocation vector, $\hat{\bgamma}$, is not the correct grouping. The point estimates of the binary vector and the rank ($\hat{\bgamma}$ and $\hat{r}$, respectively) are their corresponding maximum a posterior.

\begin{table}[t!h]
\centering
%\resizebox{\textwidth}{!}{%
{\footnotesize
\begin{tabular}{ccccccccccccccc}
\toprule
 &  &  &  &  &  & \multicolumn{4}{c}{BPRR metrics} & & \multicolumn{4}{c}{MSE $\times 10^2$} \\
$p$ & $q$ & $\qg$ & $r$ & $n$ & \qquad & $\hat{q}_\gamma$ & $\hat{r}$ & Accuracy & $F_1$ score & \qquad & BPRR & FR & RR & PRR$^*$ \\ \midrule
5 & 5 & 3 & 1 & 20 &  & 3.400 & 1.800 & 0.640 & 0.711 &  & \cellcolor{lgray}6.003 & 15.457 & 10.635 & 22.415 \\
5 & 5 & 3 & 1 & 40 &  & 3.200 & 1.300 & 0.840 & 0.864 &  & 3.969 & \cellcolor{lgray}3.807 & 5.430 & 8.189 \\
5 & 8 & 3 & 1 & 20 &  & 3.600 & 1.800 & 0.600 & 0.525 &  & \cellcolor{lgray}4.958 & 8.318 & 19.091 & 17.207 \\
% 5 & 8 & 3 & 1 & 40 &  & 4.400 & 2.100 & 0.675 & 0.667 &  & \cellcolor{lgray}3.113 & 3.853 & 6.697 & 5.482 \\
5 & 8 & 6 & 2 & 20 &  & 5.500 & 2.000 & 0.663 & 0.748 &  & \cellcolor{lgray}5.600 & 12.162 & 15.928 & 27.809 \\
5 & 8 & 6 & 4 & 20 &  & 5.800 & 3.300 & 0.725 & 0.796 &  & \cellcolor{lgray}9.892 & 21.420 & 33.727 & 29.381 \\
10 & 5 & 3 & 1 & 20 &  & 3.300 & 1.800 & 0.580 & 0.649 &  & \cellcolor{lgray}5.098 & 17.265 & 19.314 & 25.123 \\
10 & 5 & 3 & 1 & 40 &  & 3.900 & 2.500 & 0.540 & 0.671 &  & \cellcolor{lgray}2.893 & 3.751 & 4.129 & 11.351 \\
10 & 8 & 3 & 1 & 20 &  & 2.900 & 1.700 & 0.763 & 0.702 &  & \cellcolor{lgray}3.594 & 12.260 & 14.146 & 19.291 \\
10 & 8 & 6 & 2 & 20 &  & 5.600 & 3.600 & 0.600 & 0.705 &  & \cellcolor{lgray}4.546 & 19.497 & 17.502 & 27.888 \\
10 & 8 & 6 & 4 & 20 &  & 6.000 & 4.100 & 0.675 & 0.762 &  & \cellcolor{lgray}7.694 & 48.472 & 49.440 & 71.875 \\
20 & 5 & 3 & 1 & 20 &  & 2.600 & 1.400 & 0.760 & 0.794 &  & \cellcolor{lgray}4.770 & 32.919 & 34.561 & 36.030 \\
20 & 5 & 3 & 1 & 40 &  & 3.900 & 2.300 & 0.660 & 0.757 &  & \cellcolor{lgray}2.421 & 7.718 & 10.132 & 23.116 \\ \bottomrule
\end{tabular}%
}
%}
\caption{Average $\text{MSE}$ ($\times 10^2$) over 10 replicates of the listed simulation settings in columns 1 to 5, for each model: Bayesian partial reduced-rank regression (BPRR), \textit{full} low-rank regression (FR), reduced-rank regression (RR) and \textit{pre-specified allocation} partial low-rank regression (PRR$^*$). For each combination of parameters, columns 6 to 9 provide the average estimates of the number of low-rank responses, the rank of $\bC_1$, the accuracy and the $F_1$ score.}
\label{tab:sim1}
\end{table}

\begin{figure}[t!h]
    \centering
    \begin{tabular}{ccccc}   
    \vspace{-0.4cm}
    \includegraphics[width=0.18\linewidth]{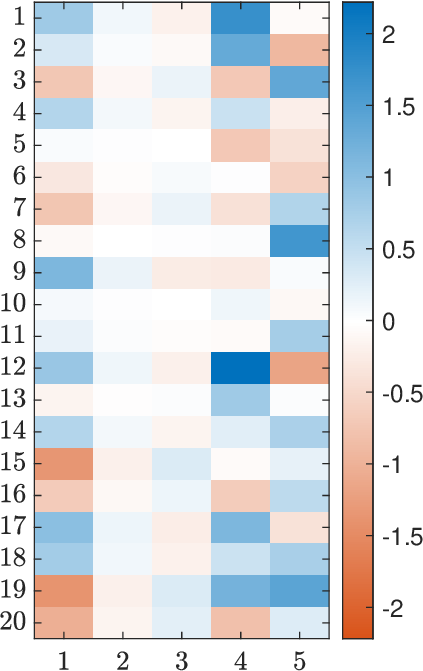} &   \includegraphics[width=0.18\linewidth]{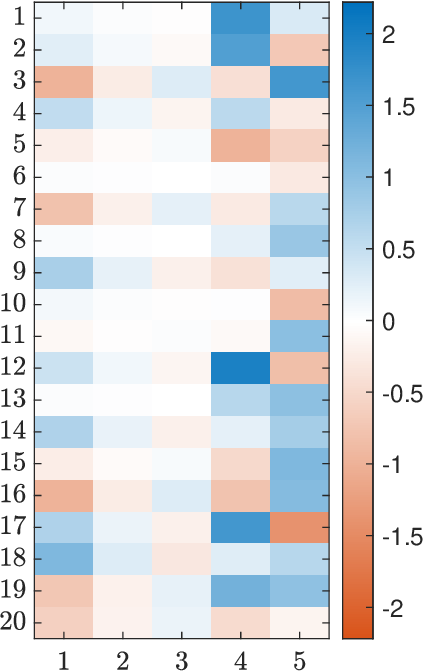}  &   \includegraphics[width=0.18\linewidth]{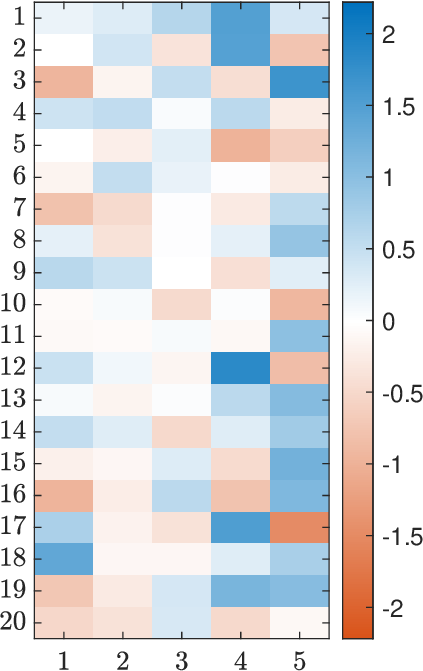}  &   \includegraphics[width=0.18\linewidth]{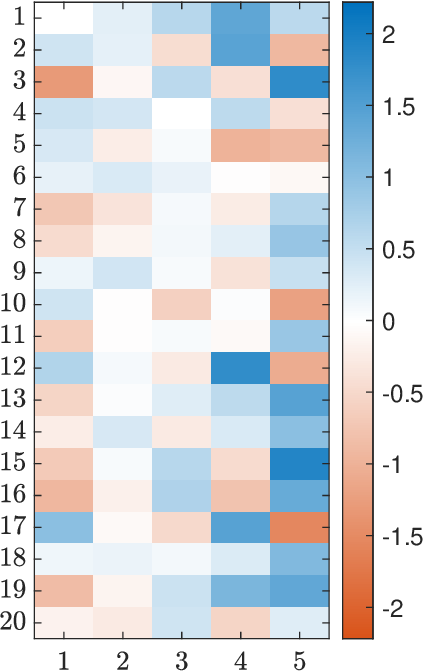}  &   \includegraphics[width=0.18\linewidth]{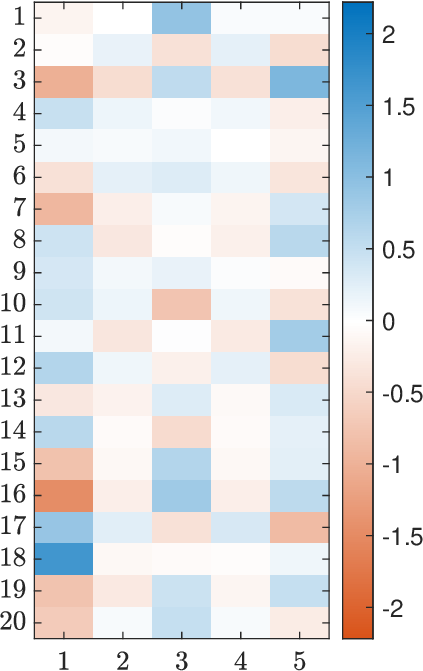} \\
    \footnotesize $\bC_0$ & \footnotesize BPRR & \footnotesize FR & \footnotesize RR & \footnotesize PRR*
    \end{tabular}
    \caption{True coefficient matrix (first left) and estimated $\bC$ matrix by each model in the simulation scenario where $p=20$, $q=5$, $\qg=3$, $r=1$ and $n=20$.}
    \label{fig:sim_Chat}
\end{figure}

Table~\ref{tab:sim1} summarises the simulation results by providing the average MSE over 10 independent experiments of each scenario. 
Our method predominantly achieves the minimum mean squared error, and increasing the number of observations results in a smaller error. 
The former result is visually explored in Fig.~\ref{fig:sim_Chat}, which examines the similarity of the coefficient matrix estimated by each of the four models with the true ordering. BPRR approximates $\bC_0$ more accurately than its competitors, and the automatic grouping choice eliminates the need for a pre-processing step to select a potentially incorrect $\bgamma$, which could result in an inferior estimation as in this case.
Additionally, we present the average estimated number of low-rank response variables with their corresponding rank estimate. As a measure of classification for the allocation parameter, we employ the accuracy and the $F_1$ score,both of which range from 0 to 1. A higher value indicates a more accurate classification of the responses.

% \gamma_0 = (1,1,0,1,0)

We observe that misspecifying a PRR regression model, either by choosing an incorrect regression model or by inaccurately determining the reduced-rank grouping, can lead to substantive performance loss, as demonstrated earlier.

\subsection{Convergence diagnostic analysis}
\label{sec:coda}

We assess the computational performance of the proposed MCMC algorithm by means of a convergence diagnostic analysis (CODA).
Specifically, for each parameter under investigation, we consider Geweke's convergence diagnostic, Heidelberger and Welch's stationarity and half-width tests \citep[see][for further details]{geweke1992evaluating,heidelberger1983simulation}.
Since the coefficient matrix $\bC$ has $pq$ entries, Table \ref{tab:coda} reports the share of entries that pass each of the aforementioned tests (i.e., $p$-value $>0.05$; ratio $< 0.10$). 
The convergence diagnostics of the rank of the low-rank matrix $\bC_1$ are equally included in Table \ref{tab:coda}. The results are satisfactory and suggest convergence of the chains for both parameters.\footnote{We performed the CODA analysis also on other independent runs of the algorithm, finding analogous results.}

\begin{table}[t!h]
\centering
{\footnotesize
%\resizebox{\textwidth}{!}{%
\begin{tabular}{cccccccccccccc}
\toprule
 &  &  &  &  &  & \multicolumn{2}{c}{Geweke test} &  & \multicolumn{2}{c}{HW Stationarity test} &  & \multicolumn{2}{c}{HW Half-width test} \\
$p$ & $q$ & $\qg$ & $r$ & $n$ &  & $r$ (p-value) & $\bC$ (share) &  & $r$ (p-value) & $\bC$ (share) &  & $r$ (ratio) & $\bC$ (share) \\ \midrule
5 & 5 & 3 & 1 & 20 &  & 0.814 & 0.920 &  & 0.376 & 1.000 &  & 0.011 & 0.800 \\
5 & 5 & 3 & 1 & 40 &  & 0.450 & 0.800 &  & 0.072 & 0.680 &  & 0.018 & 0.882 \\
%5 & 5 & 3 & 1 & 60 &  & 0.035 & 0.880 &  & 0.604 & 1.000 &  & 0.010 & 0.960 \\
5 & 8 & 3 & 1 & 20 &  & 0.193 & 0.750 &  & 0.216 & 1.000 &  & 0.037 & 0.900 \\
%5 & 8 & 3 & 1 & 40 &  & 0.026 & 0.850 &  & 0.125 & 0.975 &  & 0.015 & 0.949 \\
5 & 8 & 6 & 2 & 20 &  & 0.019 & 0.975 &  & 0.223 & 1.000 &  & 0.008 & 0.950 \\
5 & 8 & 6 & 4 & 20 &  & 0.203 & 0.975 &  & 0.290 & 0.975 &  & 0.003 & 0.949 \\
10 & 5 & 3 & 1 & 20 &  & 0.275 & 0.920 &  & 0.360 & 0.980 &  & 0.016 & 0.918 \\
10 & 5 & 3 & 1 & 40 &  & 0.795 & 1.000 &  & 0.306 & 0.980 &  & 0.015 & 1.000 \\
%10 & 5 & 3 & 1 & 60 &  & 0.472 & 1.000 &  & 0.408 & 0.980 &  & 0.012 & 0.776 \\
10 & 8 & 3 & 1 & 20 &  & 0.486 & 0.825 &  & 0.218 & 0.963 &  & 0.132 & 0.870 \\
10 & 8 & 6 & 2 & 20 &  & 0.819 & 0.975 &  & 0.550 & 1.000 &  & 0.033 & 0.813 \\
10 & 8 & 6 & 4 & 20 &  & 0.268 & 0.913 &  & 0.148 & 0.988 &  & 0.104 & 0.848 \\
20 & 5 & 3 & 1 & 20 &  & 0.011 & 0.830 &  & 0.433 & 0.990 &  & 0.004 & 0.909 \\
20 & 5 & 3 & 1 & 40 &  & 0.072 & 0.620 &  & 0.234 & 1.000 &  & 0.017 & 0.930 \\
%20 & 5 & 3 & 1 & 60 &  & 0.214 & 0.860 &  & 0.392 & 0.950 &  & 0.007 & 0.979 \\ 
\bottomrule
\end{tabular}%
}
%}
\caption{Convergence diagnostics of the rank and the coefficient matrix $\bC$:
Geweke and the Heidelberger and Welch's (HW) stationarity tests $p$-values of the rank ($>0.05$), and the rank's HW half-width test ratio ($<0.10$). Share of entries of the coefficient matrix $\bC$ that pass the Geweke test, HW stationarity and half-width tests (in column).}
\label{tab:coda}
\end{table}

\begin{figure}[t!h]
    \centering
    \begin{tabular}{cc}    
    \includegraphics[width=0.55\linewidth]{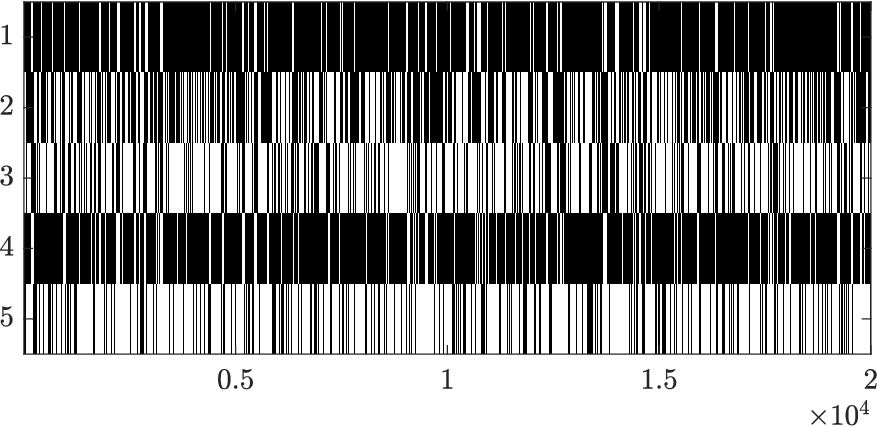} &   \includegraphics[width=0.4\linewidth]{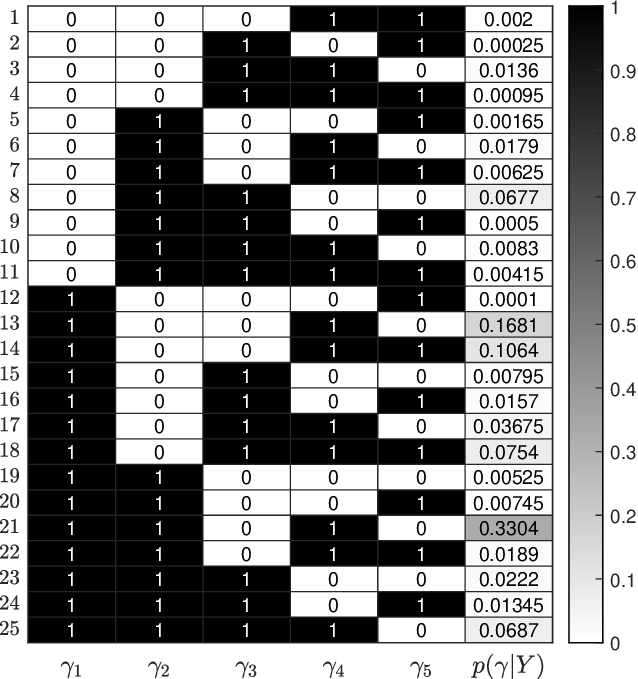}
    \end{tabular}
    \caption{Trace plot (left) and posterior distribution (right) of $\bgamma$ in the simulation scenario where $p=20$, $q=5$, $\qg=3$, $r=1$ and $n=20$.}
    \label{fig:sim_gamma_trace}
\end{figure}

Instead, as pertains to the binary allocation vector $\bgamma$, similar tests are not available. Therefore, we rely on the visual inspection of the trace plot and posterior distribution in Fig.~\ref{fig:sim_gamma_trace} to assess the convergence and mixing of the chain.
It is worth emphasising that we also performed independent runs of the algorithm (with random initialisation) and obtained similar plots. Therefore, the results in Fig.~\ref{fig:sim_gamma_trace} suggest a good exploration of the space of configurations and a good acceptance rate of the MSSS step.
In particular, the posterior distribution assigns positive mass on several configurations while displaying a unique maximum at $\hat{\bgamma} = (1,1,0,1,0)$, which corresponds to the true allocation.

% \gamma_0 = (1,1,0,1,0)

\subsection{Forecasting exercise}
\label{sec:forecasting}
As a further evaluation step, we conduct a forecasting exercise to evaluate the predictive performance of our proposed method. The dimensions of the artificially generated data for this purpose were $p=10$, $q=5$, $\qg=3$, $r=1$ and $n=60$. We utilised a window of length $40$ to make a one-step-ahead prediction for the remaining $n_{\text{test}}=20$ observations using each of the four different models: full-rank (FR), \textit{full} low-rank (RR), and \textit{pre-specified allocation} partial low-rank (PRR$^*$). A fitted matrix of responses, $\hat{\bY}$, was constructed with all $20$ predictions, and compared against the true values using the mean squared error ($\text{MSE}$) and the mean absolute error (MAE), defined as
\begin{align}
    \text{MSE} &= \sum_{i=1}^{n_{\text{test}}}{\sum_{j=1}^{q}{(\hat{y}_{ij} - y_{ij}})^2} / {(n_{\text{test}}q)}, \text{ and }
    \text{MAE} = \sum_{i=1}^{n_{\text{test}}}{\sum_{j=1}^{q}{|\hat{y}_{ij} - y_{ij}}|} / (n_{\text{test}}q).
\end{align}

The forecast error metrics of the models are presented in Table \ref{tab:fcst}, demonstrating that our model achieves superior predictive performance compared to the alternatives.
\begin{table}[t!h]
    \centering
    \resizebox{0.45\textwidth}{!}{%
    \begin{tabular}{cccccc}
    \toprule
     & \qquad & PRR & FR & RR & PRR$^*$ \\ \midrule
    MSE & \qquad & \cellcolor{lgray}1.366 & 1.503 & 1.578 & 1.962 \\
    MAE & \qquad & \cellcolor{lgray}0.910 & 0.941 & 0.982 & 1.068 \\ \bottomrule
    \end{tabular}%
    }
    \caption{Mean squared error (MSE) and mean absolute error (MAE) of the true responses versus the fitted values through a rolling forecast with the models PRR, FR, RR, and PRR$^*$.}
    \label{tab:fcst}
\end{table}

\FloatBarrier
%%%%%%%%%%%%%%%%%%%%%%%%%%%%%%%%%%%%%%%%%%%%%%%%%%%%%%%%%%%%%%%%%%%%%%%%%%

\section{An application to macroeconomic data}
\label{sec:app}

This section aims to demonstrate the usefulness of our method when applied to real-world data.
We consider quarterly macroeconomic data for the United States from 2014Q1 to 2023Q4, which were retrieved from FRED, Federal Reserve Bank of St. Louis, and the OECD, Organisation for Economic Co-operation and Development (see Section 4 of the Supplement for details).

The $q=5$ responses are the index of industrial production ($\by_1$), personal consumption of food and drinks ($\by_2$), unemployment rate ($\by_3$), volume index of imports of goods and services ($\by_4$), and volume index of exports of goods and services ($\by_5$).
The $p=5$ covariates are civilian labour force level ($\bx_1$), median weekly earnings ($\bx_2$), price index of imports of goods and services ($\bx_3$), price index of exports of goods and services ($\bx_4$), and price index of final consumption expenditure ($\bx_5$). All variables were standardised before conducting the analysis.

We investigate whether the pre-COVID period and the years following the outbreak of the COVID-19 pandemic have similar drivers.
Therefore, we consider two sub-periods by splitting the sample into the pre-COVID (2014Q1 to 2018Q4) and the (post)-COVID (2019Q1 to 2023Q4) periods, where the latter includes the outbreak of the pandemic and the subsequent recovery. Each period consists of $n=20$ quarterly observations.
Our interest lies, in particular, in investigating whether and how the estimation of the response allocation vector changes over time and quantifying the associated uncertainty.

\begin{figure}[t!h]
    \centering
    \begin{tabular}{cc}    
        \includegraphics[width=0.4\linewidth]{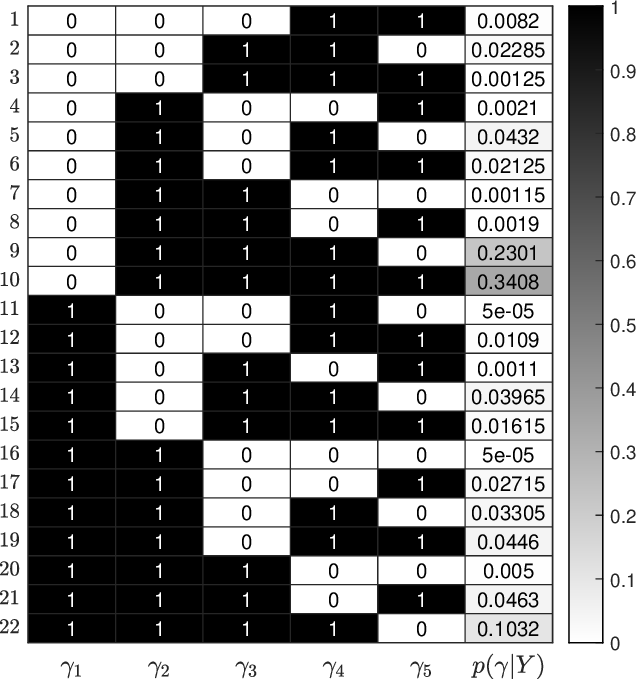} & 
        \hspace{0.25cm}
        \includegraphics[width=0.4\linewidth]{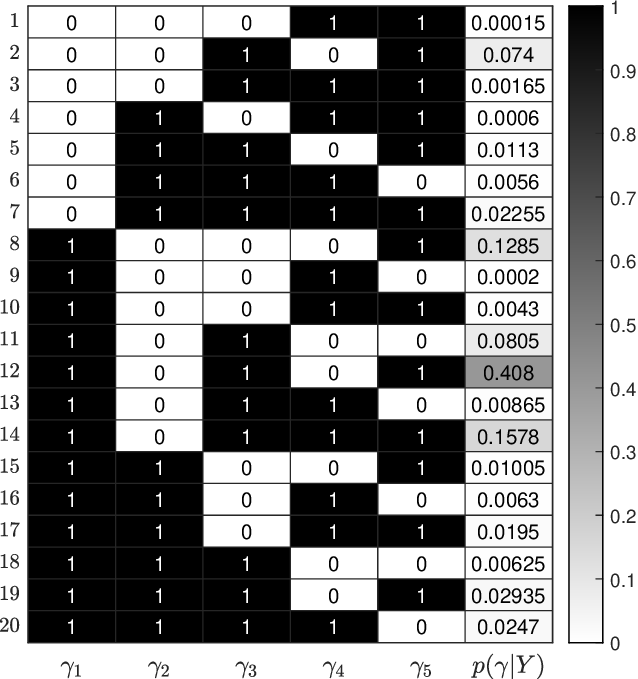}
    \end{tabular}
    \caption{Posterior distribution of the allocation vector, $\bgamma$, for the period 2014Q1-2018Q4 (left) and 2019Q1-2023Q4 (right).}
    \label{fig:app_posterior_gamma}
\end{figure}

In the first period, the estimated allocation is $\hat{\bgamma} = (0,1,1,1,1)$. However, upon inspecting the posterior distribution of $\bgamma$ in Fig.~\ref{fig:app_posterior_gamma}, it is evident that this result is highly uncertain. The allocation has posterior probability of $0.34$, followed closely by another allocation, $(0,1,1,1,0)$, at $0.23$.
In contrast, the posterior distribution of $\bgamma$ for the period including the COVID-19 pandemic exhibits a unique, clearly distinguishable mode at $\hat{\bgamma} = (1,0,1,0,1)$.
It is worth emphasising that the two periods differ not only in the point estimate of the allocation vector, $\hat{\bgamma}$, but also in the uncertainty about the estimate, which is significantly higher in the pre-COVID period. The proposed method allows us to uncover both findings directly from the data, as opposed to the traditional PRR model with \textit{a-priori fixed} allocation.

\begin{figure}[t!h]
    \centering
    \begin{tabular}{cc}    
        \includegraphics[width=0.28\linewidth]{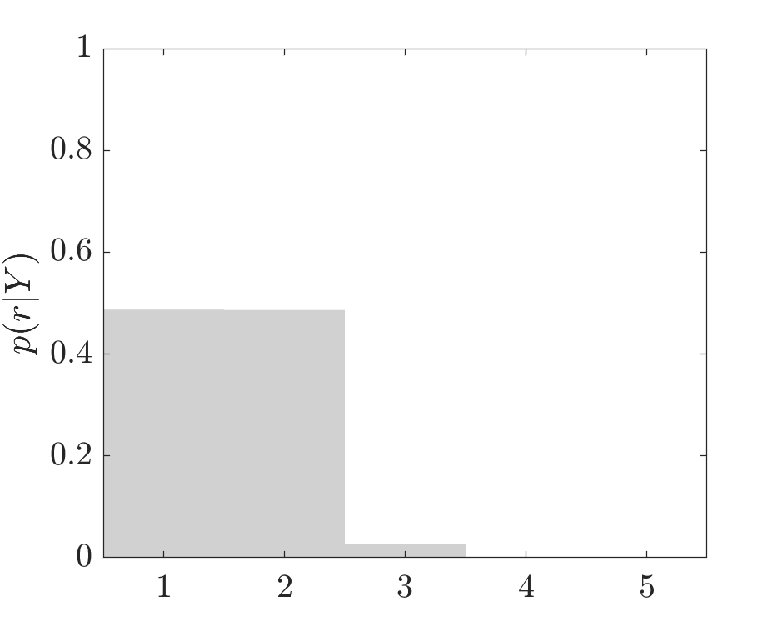} & 
        \hspace{0.10cm}
        \includegraphics[width=0.28\linewidth]{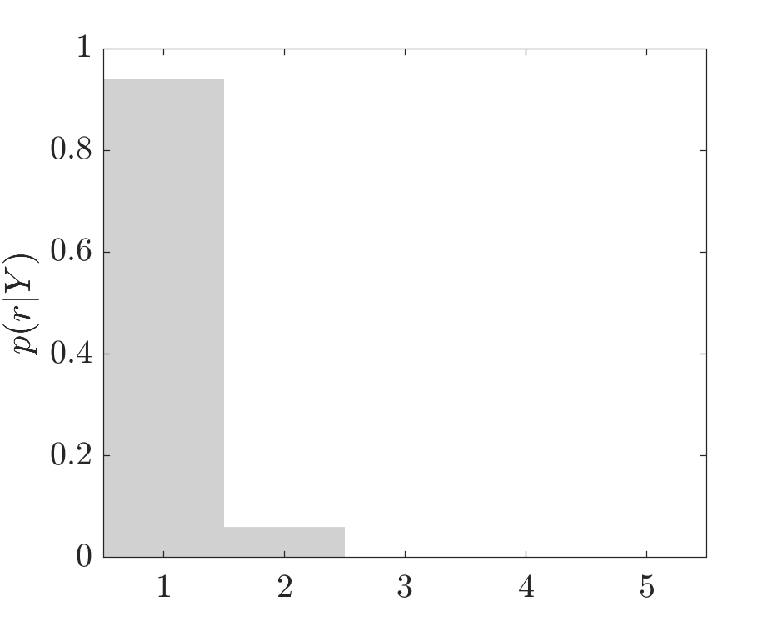}
    \end{tabular}
    \caption{Posterior distribution of the rank, $r$, for the period 2014Q1-2018Q4 (left) and 2019Q1-2023Q4 (right).}
    \label{fig:app_posterior_r}
\end{figure}

The uncertainty regarding parameter estimates for the 2014-2018 period is even more pronounced in the rank's posterior distribution, which shows nearly equal probabilities for $1$ and $2$ (Fig. \ref{fig:app_posterior_r}). 
In fact, although the point estimate of the rank is $1$, its posterior probability is $0.4873$, which is extremely close to $0.4866$ for the rank value $2$. Conversely, in the second period, we obtain a solid conclusion for the rank, as the posterior distribution has a very high mode at $1$, providing very low uncertainty about this estimate.

\begin{figure}[t!h]
    \centering
    \begin{tabular}{cc}    
        \includegraphics[width=0.23\linewidth]{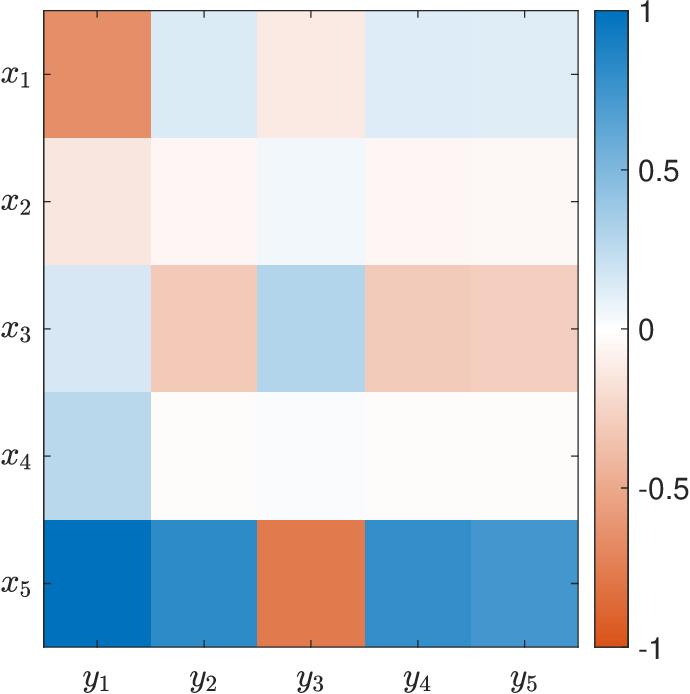} & 
        \hspace{0.5cm}
        \includegraphics[width=0.23\linewidth]{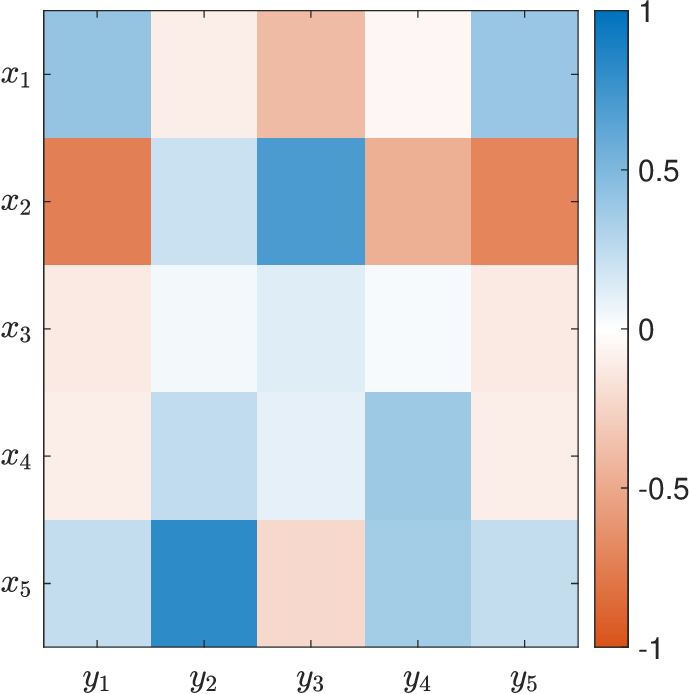}
    \end{tabular}
    \caption{Posterior mean of the coefficient matrix, $\bC$, for the period 2014Q1-2018Q4 (left) and 2019Q1-2023Q4 (right). Responses are labelled in the horizontal axis, and covariates in the vertical axis.}
    \label{fig:app_Chat}
\end{figure}

Pertaining to the regression coefficients, in Fig.~\ref{fig:app_Chat} we find evidence of a change in the relationship structure between the two periods. 
In the first period, median weekly earnings ($\bx_2$), and the price index of exports of goods and services ($\bx_4$) appear to have a negligible impact on explaining the responses of the low-rank group ($\by_2-\by_5$). This pattern changes in the subsequent quarters, where the index of industrial production ($\by_1$), the unemployment rate ($\by_3$), and the volume index of exports of goods and services ($\by_5$) exhibit a relationship with the covariates that are effectively captured by a rank-$1$ coefficient matrix. Furthermore, the weak signal of the covariates in the first period has strengthened in the second, and this estimation is more accurate in terms of the MSE compared to the other reduced-rank models presented in Section \ref{sec:sim}.\footnote{$\text{MSE}_{\text{BPRR}}=0.048$, $\text{MSE}_{\text{FR}}=0.043$,  $\text{MSE}_{\text{RR}}=0.055$, $\text{MSE}_{\text{PRR}^*}=0.049$.}

The variation between the two periods and the associated uncertainty involved suggest a research direction concerning the incorporation of time-varying parameters within a time-series framework, which could potentially facilitate the identification of structural breaks.

\FloatBarrier
%%%%%%%%%%%%%%%%%%%%%%%%%%%%%%%%%%%%%%%%%%%%%%%%%%%%%%%%%%%%%%%%%%%%%%%%%%

\section{Conclusions}
\label{sec:conc}

We have proposed a novel Bayesian approach to inference for a partial reduced rank regression model (BPRR).
To circumvent the need for transdimensional samplers, we rely on a partially collapsed Gibbs sampler, where the allocation vector and the rank parameters are drawn from their joint distribution marginalised over the coefficient matrix. Then, a Metropolis-Hasting step with local exploration is used to draw the allocation vector, reducing the unfeasible exploring of the entire space of configurations to a computationally manageable local search.

The simulation study has highlighted the good performance of the model and the proposed partially collapsed Gibbs sampler algorithm. BPRR outperforms its competitors regarding the MSE and effectively estimates the allocation vector, the rank of the reduced-rank matrix, and the regression coefficients. The MCMC convergence diagnostics support the efficacy of the algorithm.
Our approach's usefulness has also been demonstrated in real macroeconomic data, showing a significant shift in both the point estimates and posterior uncertainty about the allocation vector and the rank parameters since the outbreak of the COVID-19 pandemic.

The proposed approach can be extended in several future directions. For instance, when dealing with time series data, it would be interesting to allow the allocation vector to vary over time, that is, $\bgamma_t$.
Another point worth exploring is the design of computational tools to speed up the inferential algorithm when sampling the allocation vector \citep[e.g.,][]{geels2023taxicab}.

%%%%%%%%%%%%%%%%%%%%%%%%%%%%%%%%%%%%%%%%%%%%%%%%%%%%%%%%%%%%%%%%%%%%%%%%%%

\bibliographystyle{apalike}
\bibliography{biblio}

\end{document}